\documentclass{article}

\usepackage{microtype}
\usepackage{graphicx}
\usepackage{subfigure}
\usepackage{booktabs} 
\usepackage{mathtools}

\usepackage{hyperref}

\usepackage[accepted]{icml2019}

\icmltitlerunning{On Multi-Agent Learning in Team Sports Games}

\begin{document}

\twocolumn[
\icmltitle{On Multi-Agent Learning in Team Sports Games}



\icmlsetsymbol{equal}{*}

\begin{icmlauthorlist}
\icmlauthor{Yunqi Zhao}{EARS}
\icmlauthor{Igor Borovikov}{EARS}
\icmlauthor{Jason Rupert}{EAC}
\icmlauthor{Caedmon Somers}{EAC}
\icmlauthor{Ahmad Bierami}{EARS}
\end{icmlauthorlist}

\icmlaffiliation{EAC}{EA Sports, Electronic Arts, 4330
Sanderson Way, Burnaby, BC V5G 4X1, Canada}
\icmlaffiliation{EARS}{EA Digital Platform – Data \& AI, Electronic
Arts, Redwood City, CA 94065 USA}

\icmlcorrespondingauthor{Yunqi Zhao}{yuzhao@ea.com}
\icmlcorrespondingauthor{Igor Borovikov}{iborovikov@ea.com}
\icmlcorrespondingauthor{Ahmad Beirami}{abeirami@ea.com}

\icmlkeywords{Machine Learning, ICML}

\vskip 0.3in
]



\printAffiliationsAndNotice{}  

\begin{abstract}
In recent years, reinforcement learning has been successful in solving video games from Atari to Star Craft II. However, the end-to-end model-free   reinforcement learning (RL) is not sample efficient and requires a significant amount of computational resources to achieve superhuman level performance. Model-free RL is also unlikely to produce human-like agents for playtesting and gameplaying AI in the development cycle of complex video games. In this paper, we present a hierarchical approach to training agents with the goal of achieving human-like style and high skill level in team sports games. 
While this is still work in progress, our preliminary results show that the presented  approach holds promise for solving the posed multi-agent learning problem.
\end{abstract}

\section{Introduction}

Computer simulated environments, and particularly games, have played a central role in advancing artificial intelligence (AI). From the early days of machines playing checkers to Deep Blue, and to the most recent accomplishments of Atari bots, AlphaGo, OpenAI Dota 2 bots, and AlphaStar, artificial game agents have achieved superhuman level performance even in the most complex games. This progress is mainly due to a combination of advancements in deep learning, tree search, and reinforcement learning (RL) techniques in the past decade.

\cite{samuel-checkers} used a form of heuristic search combined with RL ideas to solve checkers.
IBM Deep Blue followed the tree search path and was the first artificial game agent who beat the chess world champion, Gary Kasparov~\cite{deep-blue}.
A decade later, Monte Carlo Tree Search (MCTS)~\cite{MCTS,UCT} was a big leap in AI to train game agents.
MCTS agents for playing Settlers of Catan were reported in~\cite{settlersszita2009monte,settlerschaslot2008monte} and shown to beat previous heuristics. \cite{Carcassonneheyden2009implementing} compares multiple approaches of agents to one another in the game Carcassonne on the two-player variant of the game and discusses variations of MCTS and Minimax search for playing the game. MCTS has also been applied to the game  of 7 Wonders~\cite{7wonders} and Ticket to Ride~\cite{MCTSTicketToRide}. 

\cite{TDGammon}, on the other hand, used TD-Lambda which is a temporal difference RL algorithm to train Backgammon agents at a superhuman level.  More recently, deep Q networks (DQNs) have emerged as a general representation learning framework from the pixels in a frame buffer combined with Q-Learning with function approximation without need for task-specific feature engineering~\cite{DQN}.\footnote{While the original DQNs worked with pixels as state space, the same idea could be applied to other cases by changing the network structure appropriately.} 
The impressive recent progress on RL to solve video games partly owes to the recent abundance of processing power and AI computing technology.\footnote{The amount of AI compute has been doubling every 3-4 months in the past few years~\cite{openai-compute}.}

DeepMind researchers remarried the two approaches by demonstrating that neural networks and their generalization properties could significantly speed up and scale MCTS. This led to AI agents that play Go at a superhuman level~\cite{alpha-go}, and solely via self-play~\cite{alpha-go-zero,alpha-zero}.
Subsequently, OpenAI researchers showed that a policy optimization approach with function approximation, called Proximal Policy Optimization (PPO)~\cite{PPO}, would lead to training agents at a superhuman level in Dota 2~\cite{openai-dota2}. The most recent progress was reported by DeepMind on StarCraft II, where AlphaStar was unveiled to play the game at a superhuman level by combining a variety of techniques including the use of attention networks~\cite{AlphaStar}.

Despite the tremendous success stories of deep RL at solving games, we believe that winning isn't everything. 
We consider the alternative problem of training human-like and believable agents that would make the video game engaging and fun for the human players.
As video games have evolved, so have the game graphics and the gameplaying AI, also referred to as game AI. 
Considering games with a limited state-action space, such as Atari games, the human-likeness and believability of AI agents would be non-issues.
Today, we have reached a point where game worlds look very realistic calling for more intelligent and realistic gameplaying agents.

While the traditional game AI solutions are already providing excellent experiences for players, it is becoming increasingly more difficult to scale those handcrafted solutions up as the game worlds are becoming larger, the content is becoming more dynamic, and the number of interacting agents is increasing. This calls for alternative approaches to train human-like and believable game AI.
We build on a variety of planning methods and machine learning techniques (including the state-of-the-art deep RL) and move away from the recent trends at training superhuman agents in solving the game AI problem~\cite{winning-isnt-everything}.

In this paper, we describe a work-in-progress hierarchical solution to a team sports video game. At a low level, the agents need to take actions that are believable and human-like whereas at a high level the agents should appear to be following a ``game plan''. While imitation learning seems apt for solving the low level problem, we propose to rely on reinforcement learning and planning to solve the high-level game strategic plan. 

The rest of the paper is organized as follows. In Section~\ref{sec:setup}, we provide the basic problem setup. In Section~\ref{sec:techniques}, we describe the solution techniques used to solve the problem. In Section~\ref{sec:MARL}, we provide a more in-depth presentation of the reinforcement learning techniques used for achieving multi-agent strategic gameplay. Finally, concluding remarks are presented in Section~\ref{sec:conclusion}.

\section{Problem Setup \& Related Work}
\label{sec:setup}

In this paper, we study a team sports video game, where the designer's goal is to train agents that exhibit strategic teamplay with a high skill level while the agents play like human players. 
Hence, the solution would entail a variety of techniques, which will be discussed in more detail in this section.

\subsection{Multi-agent learning}
Our problem naturally lends itself to the multi-agent learning (MAL) framework. 
In such a framework, iteratively optimizing for a policy could suffer from non-convergence due to the breakdown of the stationarity of the decision process and partial observability of the state space~\cite{littman94,MAL-hard}. This is because the environment for each of the agents would change whenever any other agent updates their policy, and hence independent reinforcement learning agents do not work well in practice \cite{independent-RL}.

More recently, \cite{mordatch17} proposed an actor-critic algorithm with a centralized critic during training and a decentralized actor at training and inference.
\cite{cooperative-MAL} compare policy gradient, temporal-difference
error, and actor-critic methods on cooperative deep multi-agent reinforcement learning (MARL).
See~\cite{MAL-survey,DRL-survey} for recent surveys on MAL and deep MARL advancements.

We emphasize that our problem is fundamentally simpler than the MAL framework. In contrast to the robotics problems where each agent would need to execute their own decentralized policy, in a video game all agents could be trained centrally and executed centrally as well on a single CPU. However, in the centralized treatment of the problem, in addition to the action space growing exponentially with the number of agents in the field, the chance of randomly executing a strategic play is very low, which requires collecting a huge number of state-action pairs for the agent to be able to learn such strategies if they start from random gameplay.
We will discuss some of these challenges in Section~\ref{sec:MARL}.

\subsection{Learning from demonstrations}
To ensure human-like behavior, we use human demonstrations in the training loop.
There are three general ways of using the demonstrations to train agents. Inverse reinforcement learning (IRL)
\cite{IRL, apprecnticeship-learning} would infer reward functions that promote the observed behavior in demonstrations, which can then be used in model-free RL. However, IRL is by nature an ill-posed inverse problem and tricky to solve, especially in a multi-agent framework. \cite{GAIL} proposed a direct approach to distilling a policy from the demonstrations using adversarial training, which has recently been extended to the multi-agent case~\cite{ma-airl}.

It is also possible to use demonstrations to guide RL. 
\cite{offpolicy-RL} train off-policy RL using demonstrations.
\cite{DQN} use behavioral cloning to initialize value and policy networks that would solve Go, and \cite{AlphaStar} is built on the same thought process. \cite{demonstrations-replay, magnus} use demonstrations in the replay buffer to guide the policy to a better local optimum. \cite{RL-demonstration, deepmimic} shape the reward function to promote actions that mimic the demonstrator.
\cite{bilal-safe-RL} use demonstrations to teach the policy to avoid  catastrophic events in the game of Pommerman where model-free RL fails.

\subsection{Hierarchical learning}
To manage the complexity of the posed problem (see Section~\ref{sec:techniques}), our solution involves a hierarchical approach. \cite{yisong-hierarchical} consider a hierarchical approach where the underlying low level actions are learned via RL whereas the high-level goals are picked up via IL from human demonstrations. This is in contrast to the hierarchical approach that we consider in this paper where we use IL at the low-level to achieve human-like behavior.
\cite{starcraft-hierarchy} break down the complexity of the StarCraft learning environment \cite{starcraft2} by breaking down the problem to a hierarchy of simpler learning tasks.
\cite{option-critic} apply a planning layer on top of RL where they infer the abstractions from the data as well.
Finally, \cite{feudal-RL} consider a bi-level neural network architecture where at the top level the Manager sets goals at a low
temporal resolution, and at the low level the Worker produces primitive actions conditioned on the high-level goals at a high temporal resolution.
More recently,~\cite{yisong-weak-supervision} provide a hierarchical generative model for achieving human gameplay using weak supervision.

\subsection{Human-Robot Interaction}
The human-robot interaction problem shares many similarities with the problem at hand~\cite{HRI-challenges}. However, training agents in video games is simpler in many ways. First, the agents can execute their policies centrally and there is no need for decentralized execution. Second, extracting semantic information from sensory signals such as processing images/videos and text-to-speech conversion is not needed as all of the semantic information is available from the game engine. On the other hand, many of the sample efficient learning techniques designed for training robots are applicable to training agents in team sports video games as well~\cite{HRI-imitation}. 

\section{Solution Techniques}
\label{sec:techniques}

End-to-end model-free RL requires millions of state-action pairs equivalent of many years of experience for the agent to reach human-level performance.\footnote{AlphaStar is trained using the equivalent of 60,000 years of human experience.} Applying these same techniques to modern complex games for playtesting and game AI requires obtaining and processing hundreds of years of experience, which is only feasible using significant cloud infrastructure costing millions of dollars~\cite{AlphaStar,starcraft2}. 
Hence, we move away from the end-to-end solutions in favor of  hierarchical solutions by breaking the complex problem into a hierarchy of simpler learning tasks.

We assume multiple levels of the problem abstraction in a team sports game. At the lowest level, the agent's actions and movements should resemble that of actual human players. At the highest level, the agents should learn how to follow  a (learned) high-level game plan. In the mid-level, the agents should learn to exhibit skill and to coordinate their movements with each other, e.g., to complete successful passes or to shoot toward the opponent's goal when they have a good chance of scoring.

While making RL more sample efficient is an active area of research (e.g., by curiosity-driven exploration~\cite{pathak-curiosity}), 
to apply RL to modern team sport video games or any part of the problem, we would have to shape rewards that promote a certain style or human-like behavior given human demonstrations. Reward shaping in this setup is an extremely challenging problem. 
Additionally, we also need to capture human-like cooperation/conflict in multi-agent strategic gameplay. These make reward shaping extremely challenging with mathematically vague objectives. Hence, we rely on imitation learning and behavior cloning (such as DAGGER~\cite{dagger}, learning from play (LFP)~\cite{LFP}, or GAIL~\cite{GAIL}) to achieve human-like low-level actions while we rely on RL to achieve high skill at the top level.
In this paper, we leave out the details of the imitation learning that has been used to train the low-level tactics and only focus on the mid-level strategic gameplay.

To achieve faster convergence, we rely on curriculum learning~\cite{curriculum-learning} in our training. We start training the agent to deal with easy situations more often and then make the environment more difficult. For example, the agent can learn to shoot when the opponent's net is undefended fairly quickly while it is harder to learn when to shoot when the opponent is actively defending the net. We also train the agent against simpler existing game AI agents first and then make the AI level harder once the agent has already learned the basics. Similar approaches are reported in~\cite{jiachen-curriculum} to achieve cooperation in simulating self-driving cars, and in~\cite{bilal-competition} to solve the game of Pommerman.

To deal with the MAL aspect of the problem, as the first step we train agents one at a time within the team, and let them blend into the overall strategic gameplay. Of course, there is little control gained from this process and shaping rewards is highly dependent on the status of the other agents in the environment
While we have not yet implemented the centralized training of multiple agents, this is the immediate problem we are tackling now. We will also have to solve the credit assignment in MARL for each individual agent's behavior~\cite{credit-assignment}. We remind the reader that the goal is to provide a viable approach to solving the problem with reasonable amount of computational resources.

Last but not least, we also move away from using the raw state space through the screen pixels.
 On the contrary, we provide the agent with any additional form of information that could ease training and might  otherwise be hard to infer from the screen pixels. 
 Our ultimate goal is to train human-like agents with believable behavior. Thus, so long as the agents would pass the Turing test we are not alarmed by the unfair extra information at their disposal while training. 
Furthermore, in the game development stage, 
the game itself is dynamic in the design and multiple parameters and attributes (particularly related to graphics) may change between different builds, hence it is desirable to train agents on more stable features rather than screen pixels.

\section{Strategic Gameplay via RL}
\label{sec:MARL}

\begin{figure}
    \centering
    \includegraphics[width=0.52\linewidth]{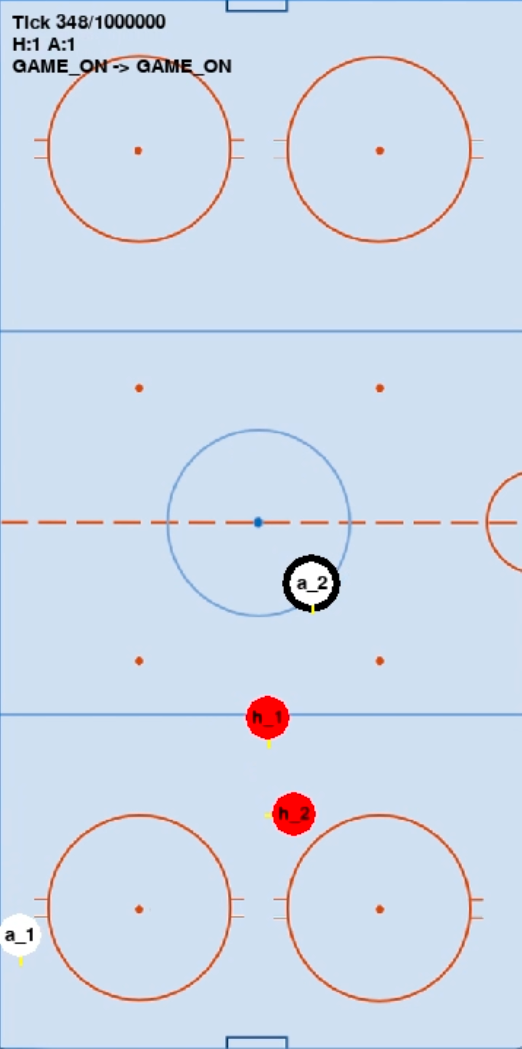}
    \caption{A screen shot of the simple team sports simulator (STS2). The red agents are home agents attempting to score at the upper end and the white agents are away agents attempting to score the lower end. The highlighted player has the possession of the ball.}\label{fig:STS2} 
  \end{figure}

Our training takes place on a  mid-level simulator, which we call simple team sports simulator (STS2).\footnote{We intend to release the STS2 gameplay environment  as an open-source package.} A screenshot of STS2 gameplay is shown in Fig.~\ref{fig:STS2}. The simulator embeds the rules of the game and the physics at a high level abstracting away the low-level tactics. The simulator supports $k$v$k$ matches for any positive integer $k$. The two teams are shown as {\em red} (home) and {\em white} (away). Each of the players can be controlled by a human, traditional game AI, or any other learned policy.
The traditional game AI consists of a handful of rules and constraints that govern the gameplay strategy of the agents.
The STS2 state space consists of the the coordinates of the players and their velocities as well as an indicator for the possession of the ball. The action space is discrete and is considered to be left, right, forward, backward, pass, and shoot. Although the player can hit two or more of the actions together we do not consider that possibility to keep the action space small for better scalability.

We currently use this mid-level simulator to inform passing and shooting decisions in the low-level imitation learning.
In the rest of this section, we report our progress toward applying deep RL in the STS2 environment to achieve multi-agent gameplay. Future work will entail a better integration between these levels of abstraction.

\subsection{Single agent in a 1v1 game}
As the simplest  first experiment, we consider training an agent that learns to play against the traditional game AI in a 1v1 match. 
We start with a sparse reward function of `+1` for scoring and `-1' for being scored against. We used DQN~\cite{DQN}, Rainbow~\cite{rainbow}, and PPO~\cite{PPO} to train agents that would replace the home team (player). DQN shows the best sign of learning useful policies after an equivalent of $\sim$5 years of human gameplay experience. The gameplay statistics of the DQN agent are reported in Table~\ref{tab:1v1-bad}. As can be seen the DQN agent was losing 1:4 to the traditional AI. Note that we randomize the orientation of the agents at the beginning of each episode, and hence, the agent encounters several easy situations with an open net for scoring. On the other hand, the agent does not learn how to play defensively when the opponent is in possession of the ball. In fact, we believe that a successful strategy for defense is more difficult to learn than that of offensive gameplay.

{
\renewcommand{\arraystretch}{1.3}
\begin{table}[h]
  \caption{DQN agent in a 1v1 match against a traditional game AI agent with a sparse `+/-1' reward for scoring. 
  }
  \label{tab:1v1-bad}
  \centering
  \footnotesize
  \begin{tabular}{l|c|c}
    Statistics & DQN Agent & Trad. Game AI \\

    \hline
    Score rate & 22\% & 78\% \\
    \hline
    Possession & 36\% & 64\% \\
  \end{tabular}
\end{table}
}

Next, we shape the rewarding mechanism with the goal of training agents that also learn how to play defensively. In addition to the `+/-1' scoring reward, we reward the agent with `+0.8' for gaining the possession of the ball and `-0.8` for losing it. 
The statistics of the DQN agent are reported in Table~\ref{tab:1v1-good}.
In this case, we observe that the DQN agent learns to play the game with an offensive style of chasing the opponent down, gaining the ball, and attempting to shoot. Its score rate as compared to the traditional game AI is 4:1, and it dominates the game.

{
\renewcommand{\arraystretch}{1.3}
\begin{table}[h]
  \caption{DQN agent in a 1v1 match against a traditional game AI agent with a sparse `+/-1' reward for scoring and a `+/0.8' reward for gaining/losing the possession of the ball. 
  }
  \label{tab:1v1-good}
  \centering
  \footnotesize
  \begin{tabular}{l|c|c}
    Statistics & DQN Agent & Trad. Game AI \\

    \hline
    Score rate & 80\% & 20\% \\
    \hline
    Possession & 65\% & 35\% \\
  \end{tabular}
\end{table}
}

We repeated this experiment using PPO and Rainbow as well. We observe that the PPO agent's policy converges quickly to a simple one. When it is in possession of the ball, it wanders around in its own half without attempting to cross the half-line or to shoot until the game times out. This happens because  the traditional game AI is programmed not to chase the opponent in their half when the opponent is in possession of the ball, and hence, the game goes on as described until timeout with no scoring on either side. PPO has clearly reached a local minimum in the space of policies, which is not unexpected as it is optimizing the policy directly.
Finally, the Rainbow agent does not learn a useful  policy for either offense or defense.

As the last 1v1 experiment, we train a PPO agent against the abovementioned DQN agent with exactly the same reward function. 
The gameplay statistics is reported in Table~\ref{tab:1v1_DQN_PPO}. We observe that the PPO agent is no longer stuck in a local optimum policy, and it is dominating the DQN agent with a score rate of 6:1. Notice that this is not a fair comparison as the DQN agent was only trained against traditional game AI agent and had not played against the PPO agent, whereas the PPO agent is directly trained against the DQN agent. While dominating the score rate, we also observe that the game is much more even in terms of the possession of the ball.
{
\renewcommand{\arraystretch}{1.3}
\begin{table}[h]
  \caption{PPO agent in a 1v1 match against a DQN agent, both with a sparse `+/-1' reward for scoring and a `+/0.8' reward for gaining/losing the possession of the ball. 
  }
  \label{tab:1v1_DQN_PPO}
  \centering
  \footnotesize
  \begin{tabular}{l|c|c}
    Statistics & PPO Agent & DQN Agent\\

    \hline
    Score rate & 86\% & 14\% \\
    \hline
    Possession & 55\% & 45\% \\
  \end{tabular}
\end{table}
}

Note that in this experiment the DQN agent is fixed, i.e., not learning, and PPO can overfit to exploit it because DQN is deterministic, easier to overfit against as an opponent. 

\subsection{Single agent in a 2v2 game}
Having gained some confidence with single agent training, as the simplest multi-agent experiment, we consider training a single agent in a 2v2 game. We let the traditional game AI be in control of the opponent players as well as the teammate player. The first experiment entails a `+/-0.8` {\em team} reward for any player in the team gaining/losing the ball in addition to the `+/-1' reward for scoring. The agent does not learn a useful defensive or offensive policy and the team loses overall.  

In the second experiment, we change the rewarding mechanism to `+/-0.8' {\em individual} reward for the agent gaining/losing the ball. 
This seems to turn the agent into an offensive player that chases the opponent down, gains the ball, and attempts to shoot.
The team statistics for this agent are shown in Table~\ref{tab:1w1v2_offensive}. We observe that the agent has learned an offensive gameplay style where it scores most of the time. 

{
\renewcommand{\arraystretch}{1.3}
\begin{table}[h]
  \caption{Offensive DQN agent in a 2v2 match against two traditional game AI agents and playing with a traditional game AI agent as teammate, with a sparse `+/-1' reward for scoring and a `+/0.8' individual reward for gaining/losing the possession of the ball. 
  }
  \label{tab:1w1v2_offensive}
  \centering
  \scriptsize
  \begin{tabular}{l|c|c|c|c}
    Statistics & DQN Agent & Teammate & Opponent 1 & Opponent 2\\

    \hline
    Score rate & 54\% & 20\% & 13\% & 13\%\\
    \hline
    Possession & 30\% & 18\%  & 26\% & 26\% \\
  \end{tabular}
\end{table}
}

While the team is winning in the previous case, we observe that the teammate is not participating much in the game with even less possession of the ball than the opponent players. Next, we explore training an agent that can assist the teammate score and possess the ball. We add another `-0.8' {\em teammate} reward, which occurs whenever the teammate loses the ball. The difference with a team reward (which resulted in an agent that did not learn defense/offense policies) here is that the agent is not getting a reward if the teammate gains puck from the opponents.
The gameplay statistics of this team are reported in Table~\ref{tab:1w1v2_defensive}. In terms of gameplay, we observe that the agent spends more time defending their own goal and passes the ball to the teammate to score when gains the possession of the ball.

{
\renewcommand{\arraystretch}{1.3}
\begin{table}[h]
  \caption{Defensive DQN agent in a 2v2 match against two traditional game AI agents and playing with a traditional game AI agent as teammate, with a sparse `+/-1' reward for scoring and a `+/0.8' individual reward for gaining/losing the possession of the ball and a `-0.8' teammate reward when the teammate loses the possession of the ball to the opponent team. 
  }
  \label{tab:1w1v2_defensive}
  \centering
  \scriptsize
  \begin{tabular}{l|c|c|c|c}
    Statistics & DQN Agent & Teammate & Opponent 1 & Opponent 2\\

    \hline
    Score rate & 20\% & 46\% & 17\% & 17\%\\
    \hline
    Possession & 36\% & 22\%  & 21\% & 21\% \\
  \end{tabular}
\end{table}
}

\subsection{Two agents trained separately in a 2v2 game}
After successful training of a single agent in a 2v2 game, we train a second agent in the home team while reusing one of the previously trained agents as the teammate. 
For this experiment, we choose the DQN agent with an offensive gameplay style from the previous set of experiments  as the teammate. This agent was described in the previous experiment.
We train another agent as the teammate using exactly the same reward function as the offensive DQN agent. 
The statistics of the gameplay for the two agents playing together against the traditional game AI agents are shown in Table~\ref{tab:2v2_offensive}. While the second agent is trained with the same reward function as the first one, it is trained in a different environment as the teammate is now the offensive DQN agent trained in the previous experiment rather than the traditional game AI agent. As can be seen, the second agent now becomes defensive and is more interested in protecting the net, gaining the possession of the ball back, and passing it to the offensive teammate.

{
\renewcommand{\arraystretch}{1.3}
\begin{table}[h]
  \caption{Two DQN agents in a 2v2 match against two traditional game AI agents, with a sparse `+/-1' reward for scoring and a `+/0.8' individual reward for gaining/losing the possession of the ball. 
  }
  \label{tab:2v2_offensive}
  \centering
  \scriptsize
  \begin{tabular}{l|c|c|c|c}
    Statistics & DQN  1& DQN  2 & Opponent 1 & Opponent 2\\

    \hline
    Score rate & 50\% & 26\% & 12\% & 12\%\\
    \hline
    Possession & 28\% & 22\%  & 25\% & 25\% \\
  \end{tabular}
\end{table}
}

As the second 2v2 experiment, we train two PPO agents in the exact same manner as we trained the DQN agents in the previous experiment. 
We observe a similar trait in the role of the agents as offensive and defensive.
Then we let the PPO team play against the DQN team.  We observe that the PPO team defeats the DQN team by a slight edge, 55:45.
While this experiment is a fair comparison between PPO and DQN, we emphasize that these teams are both trained against the traditional game AI agents and are now both playing in a new environment.
In a sense, this is measuring how generalizable the learned policy is to environments that it has not experienced before. 
The training would converge using equivalent of $\sim$5 years of human experience using DQN~\cite{DQN}. On the other hand, PPO~\cite{PPO} was an order of magnitude faster on all of the experiments reaching convergence in $\sim$6 months of human experience.

We repeated all of these experiments using Rainbow~\cite{rainbow} agents as well, and they failed all of the experiments. We suspect that the default hyperparameters in distributional RL~\cite{distributional-RL} or prioritized experience replay~\cite{prioritized-experience-replay} is  not suited to this problem, however, we are still investigating which addition in Rainbow is resulting in the failure of the algorithm in the described  team sports environment. 

\subsection{Two agents trained simultaneously in a 2v2 game} 
\label{sec:4-4}
Finally, we consider centralized training of the two home agents where a single policy controls them at the same time. We tried multiple reward functions including rewarding the team by `+1` for scoring, `-1' for being scored against, `+0.8` for gaining the possession of the ball, and `-0.8` for losing the possession of the ball.
We observed that neither algorithm learned a useful policy in this case.
We believe with a higher level planner on top of the reinforcement learning, we should be able to train the agents to exhibit teamplay but that remains for future investigation.
We are currently looking into centralized training of actor-critic methods on this environment.

\section{Concluding Remarks \& Future Work}
\label{sec:conclusion}
In this paper, we consider a team sports game. The goal is to train agents that play like humans, both in terms of tactics and  strategies. We presented a hierarchical approach to solving the problem, where the low-level problem is solved via imitation learning and the high-level problem is addressed via reinforcement learning.  We focus on strategy using a mid-level simulator, called simple team sports simulator (STS2) which we intend to release as an open-source repository. Our main takeaways are summarized below:
\begin{itemize}
    \item End-to-end model-free RL is unlikely to provide human-like and believable agent behavior, and we resort to a hierarchical approach using demonstrations to solve the problem.
    \item Sparse rewards for scoring do not provide sufficient signal for training agents, even a high level, which required us to apply more refined reward shaping. 
    \item Using proper reward shaping, we trained agents with a variety of offensive and defensive styles. In particular, we trained an agent that can assist the teammate player to achieve better scoring and ball possession.
    
    \item PPO~\cite{PPO} trained about one order of magnitude faster than DQN~\cite{DQN}, while in one occasion it got stuck in a bad local minimum.
    \item Rainbow~\cite{rainbow} failed at training agents in this environment, and we are investigating the reason this happens.
\end{itemize}

In future work, we will be working on better integrating the mid-level simulation results with the low-level imitation learned model.
We also plan to better understand and explore multi-agent credit assignment in this environment~\cite{credit-assignment}.  We also plan to investigate transfer learning for translating the policies from this environment to the actual HD game~\cite{openai-transfer}. We plan to explore further on centralized training of the multi-agent policy using QMIX~\cite{Qmix} and centralized actor-critic methods~\cite{centralized-critic}.

\section*{Acknowledgements}
The authors would like to thank Bilal Kartal (Borealis AI) and Jiachen Yang (Georgia Tech) for useful discussions and feedback. The authors are also thankful to the anonymous reviewers for their valuable feedback.

\bibliographystyle{icml2019}
\bibliography{MAS}

\end{document}